# Holographic reconstruction of the interlayer distance of bilayer two-dimensional crystal samples from their convergent beam electron diffraction patterns

*Tatiana Latychevskaia[1], Yichao Zou[2,3], Colin Robert Woods[2,4], Yi Bo Wang[2,4], Matthew Holwill[2,4], Eric Prestat[3], Sarah J. Haigh[2,3], Kostya S. Novoselov[2,4,5,6,7]*

[1]*Institute of Physics, Laboratory for ultrafast microscopy and electron scattering (LUMES), École Polytechnique Fédérale de Lausanne (EPFL) , CH-1015 Lausanne, Switzerland*

[2]*National Graphene Institute, University of Manchester, Oxford Road, Manchester, M13 9PL, UK*

[3]*School of Materials, University of Manchester, Oxford Road, Manchester, M13 9PL, UK*

[4]*School of Physics and Astronomy, University of Manchester, Oxford Road, Manchester, M13 9PL, UK*

[5]*Department of Material Science & Engineering, National University of Singapore, 117575 Singapore*

[6]*Centre for Advanced 2D Materials and Graphene Research Centre, National University of Singapore, 117546 Singapore*

[7]*Chongqing 2D Materials Institute, Liangjiang New Area, Chongqing, 400714, China*

The convergent beam electron diffraction (CBED) patterns of twisted bilayer samples exhibit interference patterns in their CBED spots. Such interference patterns can be treated as off-axis holograms and the phase of the scattered waves, meaning the interlayer distance can be reconstructed. A detailed protocol of the reconstruction procedure is provided in this study. In addition, we derive an exact formula for reconstructing the interlayer distance from the recovered phase distribution, which takes into account the different chemical compositions of the individual monolayers. It is shown that one interference fringe in a CBED spot is sufficient to reconstruct the distance between the layers, which can be practical for imaging samples with a relatively small twist angle or when probing small sample regions. The quality of the reconstructed interlayer distance is studied as a function of the twist angle. At smaller twist angles, the reconstructed interlayer distance distribution is more precise and artefact free. At larger twist angles, artefacts due to the moiré structure appear in the reconstruction. A method for the reconstruction of the average interlayer distance is presented. As for resolution, the interlayer distance can be reconstructed by the holographic approach at an accuracy of ±0.5 Å, which is a few hundred times better than the intrinsic z-resolution of diffraction limited resolution, as expressed through the spread of the measured *k*-values. Moreover, we show that holographic CBED imaging can detect variations as small as 0.1 Å in the interlayer distance, though the quantitative reconstruction of such variations suffers from large errors.

## 1. Introduction and comparison to other techniques

Convergent beam electron diffraction (CBED) [1-3] is routinely utilised for studying the parameters of thick crystals, including: thickness [4], lattice parameters [5-7] and crystallographic deformations [8, 9]. CBED performed on atomically thin two-dimensional (2D) crystals and their van der Waals structures [10, 11] produces patterns that require different interpretation than in the case of thick crystals [12-14]. Bilayer (BL) materials create a characteristic interference patterns in CBED spots, which can be treated as holograms and the phase distributions of the scattered waves, and with this, the atomic positions in the individual layers can be extracted. A particular advantage of holographic CBED is the possibility to obtain z-information from a single CBED pattern. Lateral or (x,y) atomic positions can be accessed at sub-Ångstrom resolution through a scanning procedure by electron ptychography [15]. The access to z-information possible by cross-sectional transmission electron microscopy imaging [16]. The holographic CBED approach allows access to z-atomic positions and the interlayer distance in BL systems from a single CBED pattern. In this study, we provide the theory behind and the details of the holographic reconstruction procedure applied in the holographic CBED [13]. To present a systematic study and demonstrate the performance of the technique at different parameters, we provide simulated examples.

## 2. Principle of holographic CBED reconstruction

### *2.1 Formation of CBED pattern*

The CBED arrangement in a convergent wavefront mode ($\Delta f < 0$) is shown in Fig. 1(a). A real-space distribution of a BL sample with twist angle $\varphi$ is shown in Fig. 1(b). In the virtual source plane, the Bragg diffraction peaks create virtual sources. The virtual sources of each layer are correspondingly rotated by twist angle $\varphi$, as shown in Fig. 1(c).

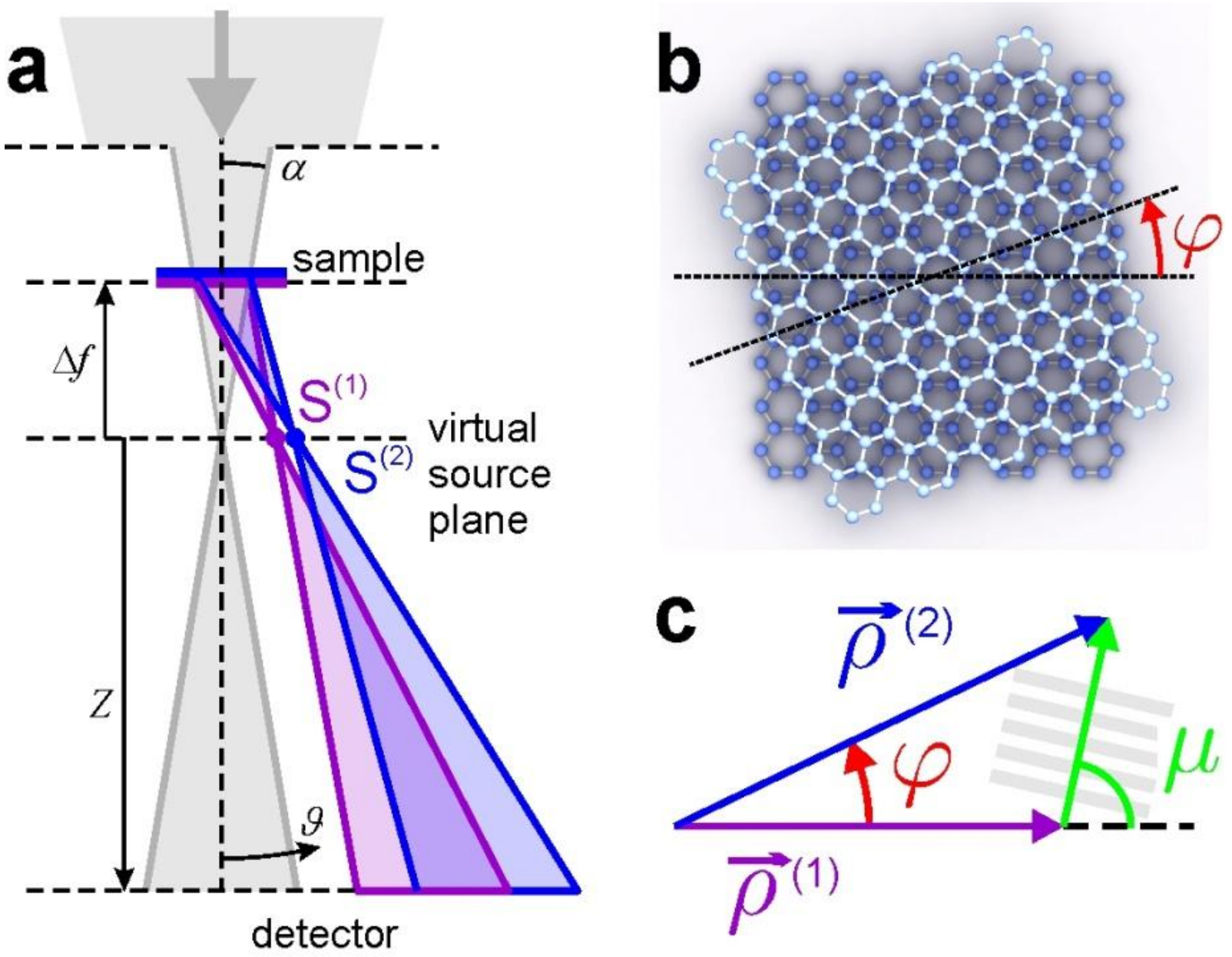

Fig. 1. CBED of bilayer sample with twist angle $\varphi$. (a) Sketch of CBED experimental arrangement. (b) Real-space distribution of the sample. (c) Arrangement of the vectors in the virtual source plane.

A convergent wavefront described by $\exp(-ikr)$ illuminates a monolayer (ML) positioned at a distance $|\Delta f|$ from the plane where the incident wavefront converges to a point (virtual source plane), as shown in Fig. 1(a); $\vec{r}$ is the coordinate in the sample plane and $k = 2\pi / \lambda$, where $\lambda$ is the wavelength . The lattice is described by a two-dimensional Dirac comb function $L(\vec{r})$. The exit wave immediately after the lattice is given by $U(\vec{r}) = \exp(-ikr)L(\vec{r})$. The wavefront propagated to the virtual source plane $(v, w)$ is calculated by the Huygens-Fresnel integral transform:

$$U_0(\vec{\rho}) = -\frac{i}{\lambda}\iint \exp(-ikr) L(\vec{r}) \frac{\exp(ik|\vec{r}-\vec{\rho}|)}{|\vec{r}-\vec{\rho}|} \,\mathrm{d}\vec{r}, \tag{1}$$

where $\vec{\rho} = (v, w)$ is the coordinate in the virtual source plane. At $r \gg \rho$, the following approximation is applied:

$$|\vec{r}-\vec{\rho}| \approx r - \frac{\vec{r}\vec{\rho}}{r} + \frac{\rho^2}{2r}$$

and Eq. (1) is re-written as:

$$U_0(\vec{\rho}) \approx -\frac{i}{\lambda}\iint L(\vec{r})\frac{1}{r}\exp\left(-ik\frac{\vec{r}\vec{\rho}}{r}\right)\exp\left(ik\frac{\rho^2}{2r}\right)\mathrm{d}\vec{r} \approx$$
$$-\frac{i}{\lambda|\Delta f|}\exp\left(ik\frac{\rho^2}{2|\Delta f|}\right)\iint L(\vec{r})\exp\left(-ik\frac{\vec{r}\vec{\rho}}{|\Delta f|}\right)\mathrm{d}\vec{r}, \quad (2)$$

where we approximate $r \approx |\Delta f|$. The integral in Eq. (2) is a Fourier transform (FT) of the lattice function $L(\vec{r})$ and the result is the reciprocal lattice defined as:

$$\tilde{L}(\vec{k}) = \iint L(\vec{r})\exp(-i\vec{k}\vec{r})\,\mathrm{d}\vec{r}.$$

Thus, for the distribution in the virtual source plane, we obtain from Eq. (2):

$$U_0(\vec{\rho}) \approx -\frac{i}{\lambda|\Delta f|}\exp\left(\frac{i\pi}{\lambda|\Delta f|}\rho^2\right)\tilde{L}\left(\frac{\vec{\rho}}{\lambda|\Delta f|}\right). \quad (3)$$

From Eq. (3), it follows that each virtual source has additional phase factor $\exp\left(\frac{i\pi}{\lambda|\Delta f|}\rho^2\right)$. Without this phase factor, the far-field diffracted wave would be given by a Dirac comb function describing the positions of individual diffraction peaks. With this phase factor, each diffraction peak is turned into a finite-sized CBED spot.

Next, we consider two layers, ML1 and ML2, with a relative twist $\varphi$ and separated by distance $d$. Each ML gives rise to a set of virtual sources and CBED spots. At relatively small $\varphi$, the CBED spots from two layers are still almost at the same positions and they overlap, creating an overlapping CBED spot. The interference pattern within each overlapping CBED spot is analogous to an interference pattern created by two waves originating from two virtual sources, as illustrated in Fig. 1(a). We assume that ML1 is shifted by $(\Delta x^{(1)}, \Delta y^{(1)})$ and ML2 is shifted by $(\Delta x^{(2)}, \Delta y^{(2)})$ relatively to a centred lattice (a lattice in which one of its hexagons is centred at the origin of the $(x, y)$ plane). ML1 is positioned at a distance $|\Delta f|$ from the virtual source plane, ML2 is positioned at a distance $|\Delta f + d|$ from the virtual source plane. The two corresponding wavefront distributions in the virtual source $(v, w)$ plane are given by Eq. (3):

$$U_0^{(1)}(\vec{\rho}) \propto \frac{1}{|\Delta f|}\exp\left(\frac{\pi i}{\lambda|\Delta f|}\left(\rho^{(1)}\right)^2\right)\delta\left(\vec{\rho}-\vec{\rho}_m^{(1)}\right),$$
$$U_0^{(2)}(\vec{\rho}) \propto \frac{1}{|\Delta f + d|}\exp\left(\frac{\pi i}{\lambda|\Delta f + d|}\left(\rho^{(2)}\right)^2\right)\delta\left(\vec{\rho}-\vec{\rho}_m^{(2)}\right), \quad (4)$$

where $\vec{\rho}_m^{(1)}$ and $\vec{\rho}_m^{(2)}$ are the positions of the virtual sources in the $(v,w)$-plane and $m$ ($m=1...6$) is a CBED spot number. In Eq. (4), we neglect the secondary scattering of the electron wave on the second layer assuming that the second layer is also illuminated with a plane (convergent) wave. Each virtual source creates a divergent spherical wave described by $\exp\left(ik\left|\vec{R}-\vec{\rho}_m\right|\right)$, $\vec{R}$ is the coordinate in the detector plane.

The interference pattern within an overlapping CBED spot is described as:

$$I_m(R)\propto$$
$$\propto\left|\exp\left(\frac{i\pi}{\lambda|\Delta f|}\left(\rho^{(1)}\right)^2\right)\exp\left(ik\left|\vec{R}-\vec{\rho}_m^{(1)}\right|+i\gamma_m^{(1)}\right)+\exp\left(\frac{i\pi}{\lambda|\Delta f+d|}\left(\rho^{(2)}\right)^2\right)\exp\left(ik\left|\vec{R}-\vec{\rho}_m^{(2)}\right|+i\gamma_m^{(2)}\right)\right|^2, \quad (5)$$

where $\gamma_m^{(1)}$ and $\gamma_m^{(2)}$ are the constant phases of the virtual sources defined as follows. When an object is shifted by $\left(\Delta x^{(i)},\Delta y^{(i)}\right)$, $i=1,2$, its inverse Fourier transform gains an addition linear phase shift given by the additional factor:

$$\exp\left[\frac{2\pi i}{\lambda|\Delta f|}\left(v\Delta x^{(i)}+w\Delta y^{(i)}\right)\right].$$

A virtual source at $\vec{\rho}_m^{(i)}=\left(v_m^{(i)},w_m^{(i)}\right)$ gains a constant phase shift:

$$\exp\left[\frac{2\pi i}{\lambda|\Delta f|}\left(v_m^{(i)}\Delta x^{(i)}+w_m^{(i)}\Delta y^{(i)}\right)\right]=\exp\left(i\gamma_m^{(i)}\right),$$

where we introduce:

$$\gamma_m^{(i)}=\frac{2\pi}{\lambda|\Delta f|}\left(v_m^{(i)}\Delta x^{(i)}+w_m^{(i)}\Delta y^{(i)}\right). \quad (6)$$

Note that $\gamma_m^{(i)}$ has the opposite sign for opposite CBED spots, because of the opposite signs of the coordinates $\left(v_m^{(i)},w_m^{(i)}\right)$. This feature is important for reconstruction of *d* by averaging the reconstructed phase over all six CBED spots, because the averaging removes $\gamma_m^{(i)}$ terms. In the following text, we omit the subscript $m$.

The following approximation can be applied:

$$\left|\vec{R}-\vec{\rho}\right|=\sqrt{R^2-2\vec{R}\vec{\rho}+\rho^2}\approx R-\vec{R}\vec{\rho}/R=R-\vec{K}\vec{\rho}$$

where $\vec{K}=k\frac{\vec{R}}{R}$, and we can rewrite Eq. (5) as:

$$I(R) \propto$$

$$\propto \left| \exp\left[ \frac{ik}{2|\Delta f|} \left(\rho^{(1)}\right)^2 \right] \exp\left( ik\left|\vec{R}-\vec{\rho}^{(1)}\right| + i\gamma^{(1)} \right) + \exp\left[ \frac{ik}{2|\Delta f + d|} \left(\rho^{(2)}\right)^2 \right] \exp\left( ik\left|\vec{R}-\vec{\rho}^{(2)}\right| + i\gamma^{(2)} \right) \right|^2 =$$

$$= \left| \exp\left[ -ik\frac{\vec{R}\vec{\rho}^{(1)}}{R} + ik\frac{\left(\rho^{(1)}\right)^2}{2R} + ik\frac{\left(\rho^{(1)}\right)^2}{2|\Delta f|} + i\gamma^{(1)} \right] + \exp\left[ -ik\frac{\vec{R}\vec{\rho}^{(2)}}{R} + ik\frac{\left(\rho^{(2)}\right)^2}{2R} + ik\frac{\left(\rho^{(2)}\right)^2}{2|\Delta f + d|} + i\gamma^{(2)} \right] \right|^2 =$$

$$= \left| \exp\left[ ik\frac{\vec{R}\vec{\rho}^{(1)}}{R} - ik\frac{\left(\rho^{(1)}\right)^2}{2R} - ik\frac{\left(\rho^{(1)}\right)^2}{2|\Delta f|} - i\gamma^{(1)} \right] + \exp\left[ ik\frac{\vec{R}\vec{\rho}^{(2)}}{R} - ik\frac{\left(\rho^{(2)}\right)^2}{2R} - ik\frac{\left(\rho^{(2)}\right)^2}{2|\Delta f + d|} - i\gamma^{(2)} \right] \right|^2 =$$

$$= 2 + 2\cos\left\{ k\frac{\vec{R}}{R}\left(\vec{\rho}^{(1)} - \vec{\rho}^{(2)}\right) - \frac{k}{2R}\left[\left(\rho^{(1)}\right)^2 - \left(\rho^{(2)}\right)^2\right] + \frac{k}{2}\left[ \frac{\left(\rho^{(2)}\right)^2}{|\Delta f + d|} - \frac{\left(\rho^{(1)}\right)^2}{|\Delta f|} \right] - \Delta\gamma \right\} \approx$$

$$2 + 2\cos\left\{ \vec{K}\Delta\vec{\rho} + \frac{\pi}{\lambda}\left[ \frac{\left(\rho^{(2)}\right)^2}{|\Delta f + d|} - \frac{\left(\rho^{(1)}\right)^2}{|\Delta f|} \right] - \Delta\gamma \right\}, \qquad (7)$$

where $\Delta\vec{\rho} = \vec{\rho}^{(1)} - \vec{\rho}^{(2)}$ and $\Delta\gamma = \gamma^{(1)} - \gamma^{(2)}$. Equation (7) is a general formula that describes the interference pattern in an overlapping CBED spot of a bilayer. The first term in the argument of cosine describes the interference fringes and therefore provides the positions of the sidebands in the Fourier spectrum of the CBED spot. The second term in the argument of cosine is a constant offset which together with the first term allows recovery of the interlayer distance *d*. The third term $\Delta\gamma$ depends on the local stacking of the layers (for example, AA or AB stacking) and defines the position of the centre of the interference pattern. If the local stacking under the centre of the electron beam is AA, then $\Delta\gamma = 0$ and the interference pattern within an overlapping CBED spot has a maximum at the centre of the CBED spot.

## *2.2 Extracting the interlayer distance from the inference pattern*

The information regarding the interlayer distance is enclosed in the first and second terms of the argument of the cosine in Eq. (7). We consider the interference pattern in a first-order CBED spot. For a BL, we introduce the virtual sources coordinates as:

$$\vec{\rho}^{(1)} = \left(|\Delta f|\tan\vartheta^{(1)}, 0\right)$$

$$\vec{\rho}^{(2)} = \left(|\Delta f + d|\tan\vartheta^{(2)}\cos\varphi, |\Delta f + d|\tan\vartheta^{(2)}\sin\varphi\right)$$

where $\vartheta^{(1)}$ and $\vartheta^{(2)}$ are the diffraction angles corresponding to ML1 and ML2, as shown in Fig. 1(c). The $\vec{K}$ coordinate corresponds to the centre of the overlapping CBED spot and is given by the average of the first-order diffraction coordinates of ML1 and ML2:

$$\vec{K}_{\text{avg}} = \frac{\vec{K}^{(1)} + \vec{K}^{(2)}}{2},$$

where:

$$\vec{K}^{(1)} = \frac{2\pi}{\lambda}\left(\sin\vartheta^{(1)}, 0\right)$$
$$\vec{K}^{(2)} = \frac{2\pi}{\lambda}\left(\sin\vartheta^{(2)}\cos\varphi, \sin\vartheta^{(2)}\sin\varphi\right).$$

The first term in cosine argument of Eq. (7) gives:

$$\vec{K}_{\text{avg}}\Delta\vec{\rho} =$$
$$\frac{\pi}{\lambda}\left(|\Delta f|\tan^2\vartheta^{(1)} + |\Delta f|\tan\vartheta^{(2)}\cos\varphi\tan\vartheta^{(1)} - |\Delta f + d|\tan\vartheta^{(2)}\cos\varphi\tan\vartheta^{(1)} - |\Delta f + d|\tan^2\vartheta^{(2)}\right),$$

where we assumed that $\sin\vartheta^{(i)} \approx \tan\vartheta^{(i)}$. The second term in the cosine argument of Eq. (7) gives:

$$\frac{\pi}{\lambda}\left[\frac{\left(\rho^{(2)}\right)^2}{|\Delta f + d|} - \frac{\left(\rho^{(1)}\right)^2}{|\Delta f|}\right] = \frac{\pi}{\lambda}\left[\frac{|\Delta f + d|^2\tan^2\vartheta^{(2)}}{|\Delta f + d|} - \frac{|\Delta f|^2\tan^2\vartheta^{(1)}}{|\Delta f|}\right] = \frac{\pi}{\lambda}\left[|\Delta f + d|\tan^2\vartheta^{(2)} - |\Delta f|\tan^2\vartheta^{(1)}\right]$$

The sum of the two terms gives:

$$\chi = \vec{K}_{\text{avg}}\Delta\vec{\rho} + \frac{\pi}{\lambda}\left[\frac{\left(\rho^{(2)}\right)^2}{|\Delta f + d|} - \frac{\left(\rho^{(1)}\right)^2}{|\Delta f|}\right] = \frac{\pi d}{\lambda}\tan\vartheta^{(2)}\tan\vartheta^{(1)}\cos\varphi \quad (8)$$

Thus, the distribution of the interlayer distance over the probed region $d(x, y)$ can be extracted from the sum of the two first terms $\chi$. Equation (8) implies that the interlayer distance can be determined even if the MLs are of different chemical compositions.

For identical MLs $\vartheta^{(1)} = \vartheta^{(2)} = \vartheta$ and at small twist angles, we obtain:

$$\chi = \frac{\pi d}{\lambda}\tan^2\vartheta\cos\varphi \approx \frac{\pi d\lambda}{a^2},$$

where *a* is the lattice constant. For non-identical MLs provided $\vartheta^{(1)} \approx \vartheta^{(2)}$, the extracted $d$ will be different from the real $d_0$ by the factor $d = d_0\left(1 - \Delta\vartheta / \vartheta^{(1)}\right)$, where $\Delta\vartheta = \vartheta^{(2)} - \vartheta^{(1)}$. For example, for graphene and boron nitride (BN) layers, this error factor amounts to $\left(1 - \Delta\vartheta / \vartheta^{(1)}\right) = 0.98$.

## 3. Reconstruction of CBED spots as off-axis holograms

In this section, we provide a step-by-step protocol for the holographic reconstruction.

### 3.1 Positions of the CBED spots

The centres of the CBED spots in the CBED pattern, as Bragg diffraction peaks, are given by the period of the lattice $d_i$:

$$K^{(i)} = \frac{2\pi}{d_i}, i = 1, 2.$$

The positions of the CBED spots are theoretically calculated from the period and the rotation of the lattice. For a hexagonal lattice, the first-order CBED spots are given by the lattice period $d_i = \frac{\sqrt{3}}{2} a_i$, where $a_i$ is the lattice constant, with values of 0.246 and 0.250 nm for graphene and BN, respectively.

### 3.2 Selecting the centre of the overlapping CBED spots

For a twisted BL sample, the CBED pattern consists of two overlapping CBED patterns, each from individual ML. CBED spots overlap in pairs, as shown in Fig. 2. For holographic reconstruction, the centre of an overlapping spot is calculated as an arithmetic average of the centres of the individual CBED spots:

$$\vec{K}_{\mathrm{avg}} = \frac{\vec{K}^{(1)} + \vec{K}^{(2)}}{2},$$

where $\vec{K}^{(1)}$ and $\vec{K}^{(2)}$ are the coordinates of the centres of the CBED spots from the individual MLs. A square region is selected with the centre at $\vec{K}_{\mathrm{avg}}$ at an overlapping CBED spot (as indicated by the red dot in Fig. 2(b)).

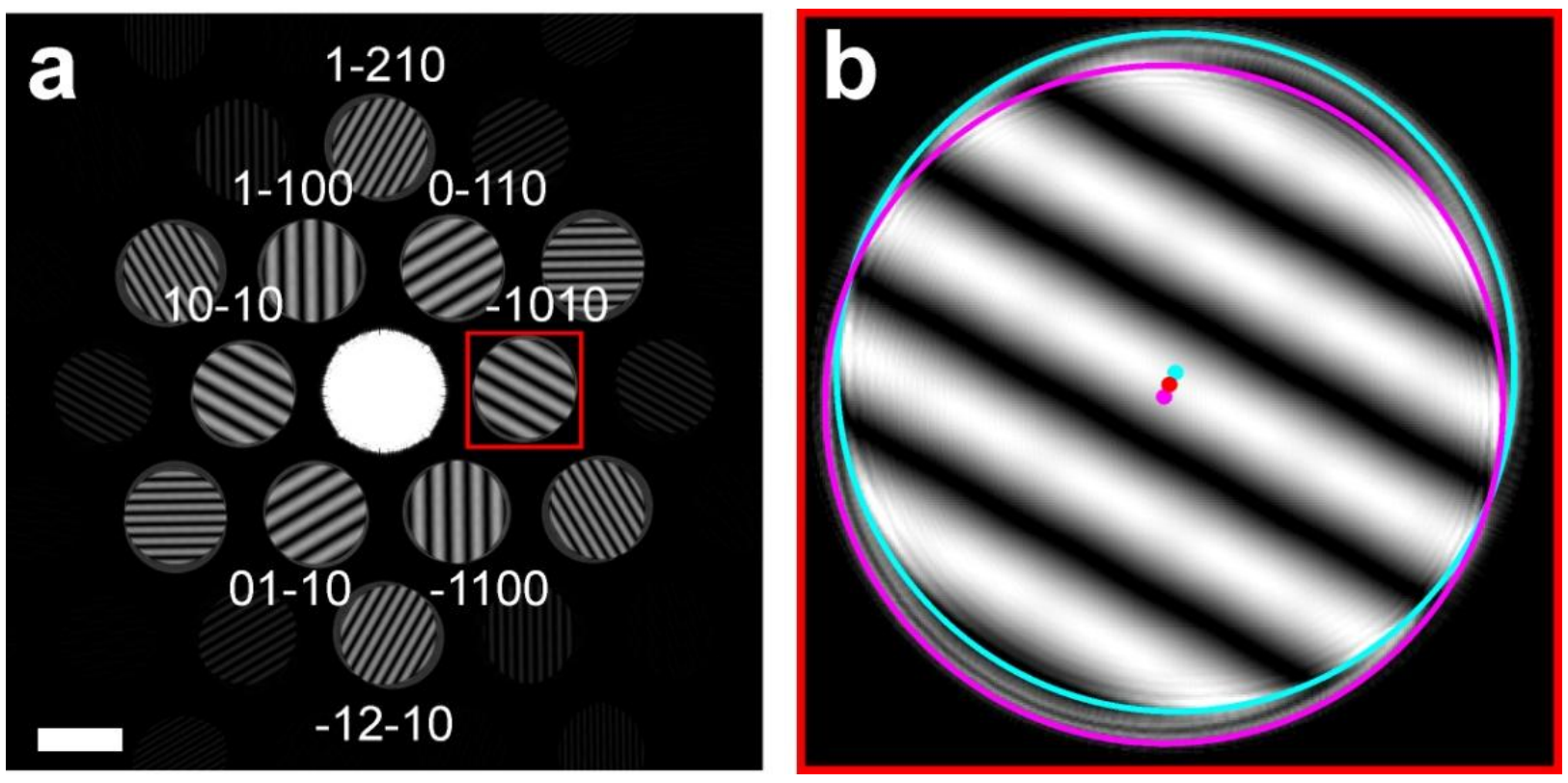

Fig. 2. Selection of an overlapping CBED spot for holographic reconstruction. (a) Simulated CBED pattern of graphene/hBN with a twist angle of 2°. (b) Magnified selected overlapping CBED spot (-1010) (shown in the red square in (a)). The CBED spot from the graphene layer and its centre are shown in cyan. The CBED spot from the BN layer and its centre are shown in magenta. The centre of the overlapping CBED spot is shown as the red dot.

### *3.3 Reconstructing the CBED spot as an off-axis hologram*

The intensity distribution in a selected CBED spot can be re-written as:

$$I\left(\vec{K}'\right) \propto 2\cos\left(\vec{K}'\Delta\vec{\rho} + \chi - \Delta\gamma\right),$$

where $\vec{K}'$ is the running coordinate counted from the centre of the selected overlapping CBED spot and $\chi$ is introduced in Eq. (8). The intensity distribution $I\left(\vec{K}'\right)$ is treated as an off-axis hologram and a conventional protocol of an off-axis hologram reconstruction is applied [17, 18]. The 2D FT of $I\left(\vec{K}'\right)$ gives:

$$\mathrm{FT}\left[I\left(\vec{K}'\right)\right] \propto \iint \exp\left[-i\left(K'_x v + K'_y w\right)\right] \times$$
$$\left\{\exp\left[i\left(K'_x \Delta\rho_x + K'_y \Delta\rho_y + \chi\right)\right] + \exp\left[-i\left(K'_x \Delta\rho_x + K'_y \Delta\rho_y + \chi\right)\right]\right\} \mathrm{d}K'_x \, \mathrm{d}K'_y =$$
$$\exp\left(i\chi\right)\delta\left(v - \Delta\rho_x, w - \Delta\rho_y\right) + \exp\left(-i\chi\right)\delta\left(v + \Delta\rho_x, w + \Delta\rho_y\right)$$

The right sideband corresponds to the term $\exp\left(i\chi\right)\delta\left(v - \Delta\rho_x, w - \Delta\rho_y\right)$ and the left sideband corresponds to $\exp\left(-i\chi\right)\delta\left(v + \Delta\rho_x, w + \Delta\rho_y\right)$. The right sideband is selected and shifted to the centre. The inverse 2D FT of the resulting distribution gives the reconstructed complex-valued distribution $\exp\left(i\chi\right)$. In principle, either the left or right sideband can be selected since they both carry the same information. When the left sideband is selected, $\exp\left(-i\chi\right)$ is reconstructed and the sign of the reconstructed distribution should be flipped.

The reconstruction steps are illustrated in Fig. 3 and discussed below:

(1) The 2D Fourier spectrum of the hologram is calculated by 2D FT. In the spectrum, one zero-order and two sidebands are observed (Figs. 3(a) and (b)).

(2) The right sideband is selected and the zero-order and left sideband are set to zero (Fig. 3(b)).

(3) The whole spectrum is shifted so that the selected sideband is in the centre (Fig. 3(b)).

(4) The inverse 2D FT of the resulting distribution gives a complex-valued reconstruction, where the amplitude and the phase distributions can be extracted, (Fig. 3(c)).

In steps (3) and (4), the right sideband is selected at the position defined by $\mu$ and *T*, as explained in the next section.

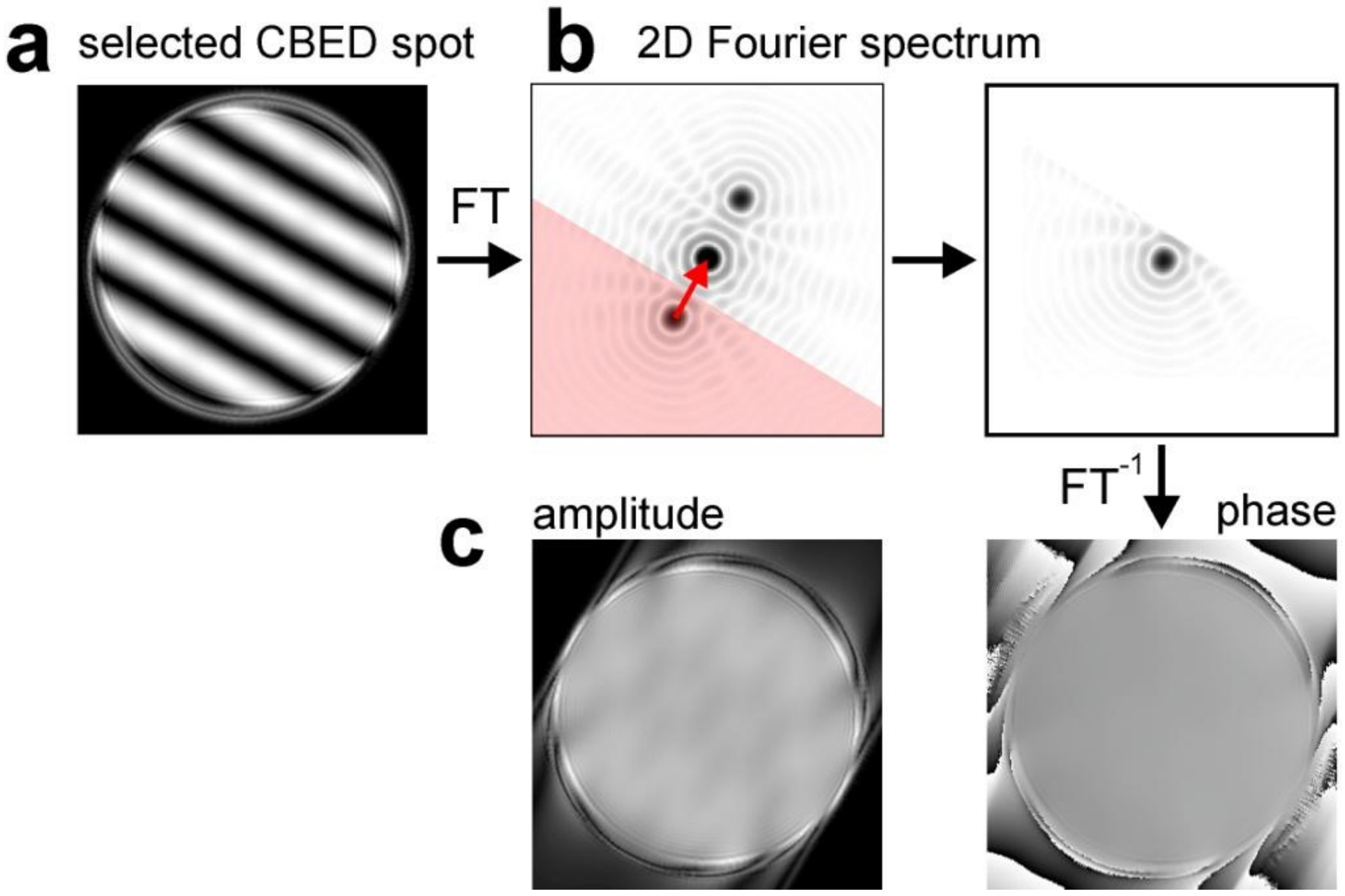

Fig. 3. Reconstruction of a CBED spot as an off-axis hologram. (a) CBED spot (-1010) (shown in Fig. 2(a)). (b) Amplitude of its Fourier spectrum. The area marked by red is selected and shifted so that the sideband peak is in the centre. (c) Inverse 2D FT gives the complex-valued distribution, where the amplitude and phase are extracted.

## *3.4 Period and tilt of fringes*

The period of the interference fringes is given by:

$$T = \frac{2\pi}{\Delta\rho}, \tag{9}$$

where $\Delta\rho$ is given by:

$$(\Delta\rho)^2 = \left(\rho^{(1)}\right)^2 + \left(\rho^{(2)}\right)^2 - 2\rho^{(1)}\rho^{(2)}\cos\varphi.$$

The tilt of the interference fringes $\mu$ can be found from the geometrical arrangement of the vectors in the virtual source plane (Fig. 1(c)):

$$\tan\mu = \frac{\rho^{(2)}\sin\varphi}{\rho^{(2)}\cos\varphi - \rho^{(1)}}. \tag{10}$$

$\Delta f$ is approximately known from experiment and $d$ is typically unknown. From Eqs. (9) and (10), $T$ and $\mu$ can be evaluated from the known $\vartheta^{(1)}$, $\vartheta^{(2)}$, $\varphi$ and $\Delta f$, assuming the distance between the layers $d = 0$. The period of fringes $T$ and the tilt of fringes $\mu$ allow for calculation of the exact positions of the sidebands in the Fourier spectrum.

### *3.5 Re-positioning of reconstructed distribution*

From the complex-valued distribution reconstructed in the previous step, the distribution corresponding to one of the layers, for example ML1, needs to be selected. ML1 CBED spots are positioned at $\Delta\vec{K}^{(1)} = \vec{K}_{\mathrm{avg}} - \vec{K}^{(1)}$ offset from the centres of the overlapping spots, as illustrated in Fig. 4. Therefore, the reconstructed distributions need to be re-positioned to correspond to the position of the CBED spots of ML1. This is achieved by shifting the reconstructed distributions by $\Delta\vec{K}^{(1)} = \vec{K}_{\mathrm{avg}} - \vec{K}^{(1)}$.

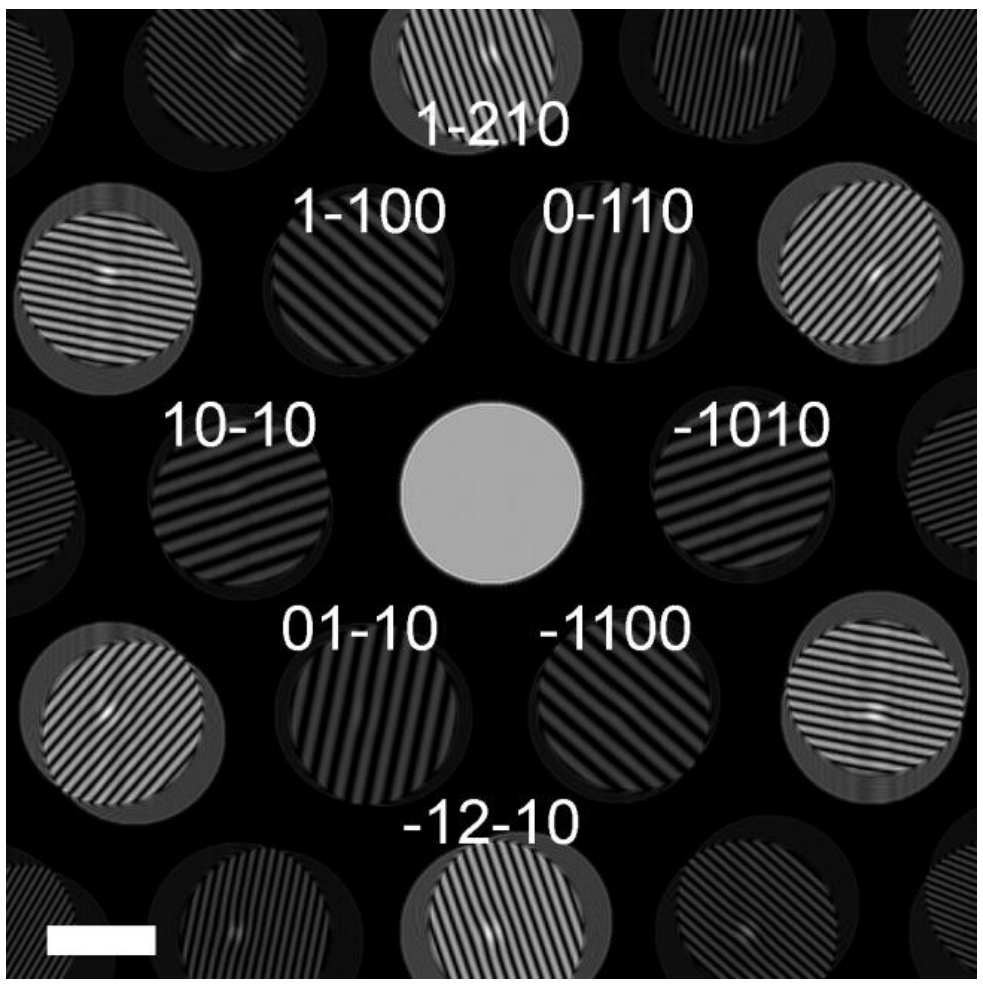

Fig. 4. Simulated CBED patterns for graphene-BN BL where the graphene lattice is deformed. Atoms in graphene ML are displaced as follows. Atoms positioned at $x > 0$ are displaced by $\Delta x = -10$ pm, and the atomic $z$-positions are shifted by $\Delta z = -A\exp\left(-\frac{x^2 + y^2}{2\sigma^2}\right)$, $A = 2$ nm and $\sigma = 2$ nm. The deformations in the graphene ML help to show that the graphene CBED spots are positioned at $\Delta\vec{K}^{(1)} = \vec{K}_{\mathrm{avg}} - \vec{K}^{(1)}$ offset from the centres of the overlapping spots. For these simulations $\Delta f = -2$ μm, the distance between the layers is 3.35 Å and the imaged area is 28 nm in diameter. The scale bar corresponds to 2 nm$^{-1}$.

### *3.6 Averaging*

After the amplitude and phase distributions for each CBED spot are reconstructed, only the phase distributions are considered since only these distributions carry the information regarding the atomic positions. The individual reconstructed phase distributions are averaged, that is, all six distributions are added together and divided by six. As mentioned above by Eq. (6), averaging also eliminates $\Delta\gamma$.

### *3.7 Extracting the interlayer distance*

The interlayer distance is obtained from the reconstructed averaged phase distribution by applying Eq. (8).

## 4. Effect of various parameters

### *4.1 Number of fringes*

The number of fringes in an overlapping CBED spot is given by the twist angle and the size of the probed region, which in turn is given by the defocus value. In this section, we show that even one interference fringe (one period) is sufficient to reconstruct the phase shift.

Figure 5 shows simulated CBED patterns for BL graphene (BLG) with the interlayer distance of 10 Å, at defocus -2 µm and three different twist angles of 0.5°, 2° and 4°. While CBED patterns with the twist angles 2° and 4° exhibit a few interference fringes in their CBED spots, the CBED pattern with the twist angle of 0.5° exhibits only one period of the interference fringes. As a consequence, the zero-order and sideband in the spectrum of this CBED pattern are not well resolved and cannot be clearly separated one from another, as shown in the inset in Fig. 5(a). However, applying the reconstruction procedure described above still provides correct reconstruction and the interlayer distance, as shown in Fig. 6.

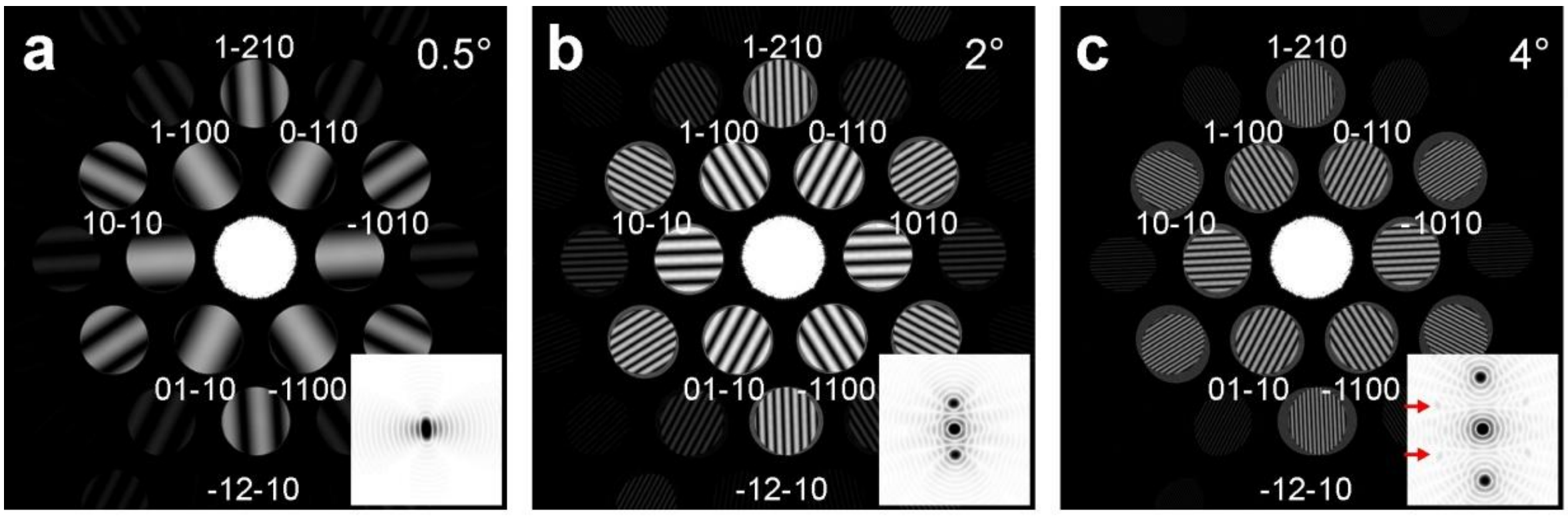

Fig. 5. Simulated CBED patterns for BLG with the interlayer distance of 10 Å, defocus of -2 µm and three different twist angles: (a) 0.5°, (b) 2° and (c) 4°. The corresponding Fourier spectra of (-1010) spots are shown in the corners.

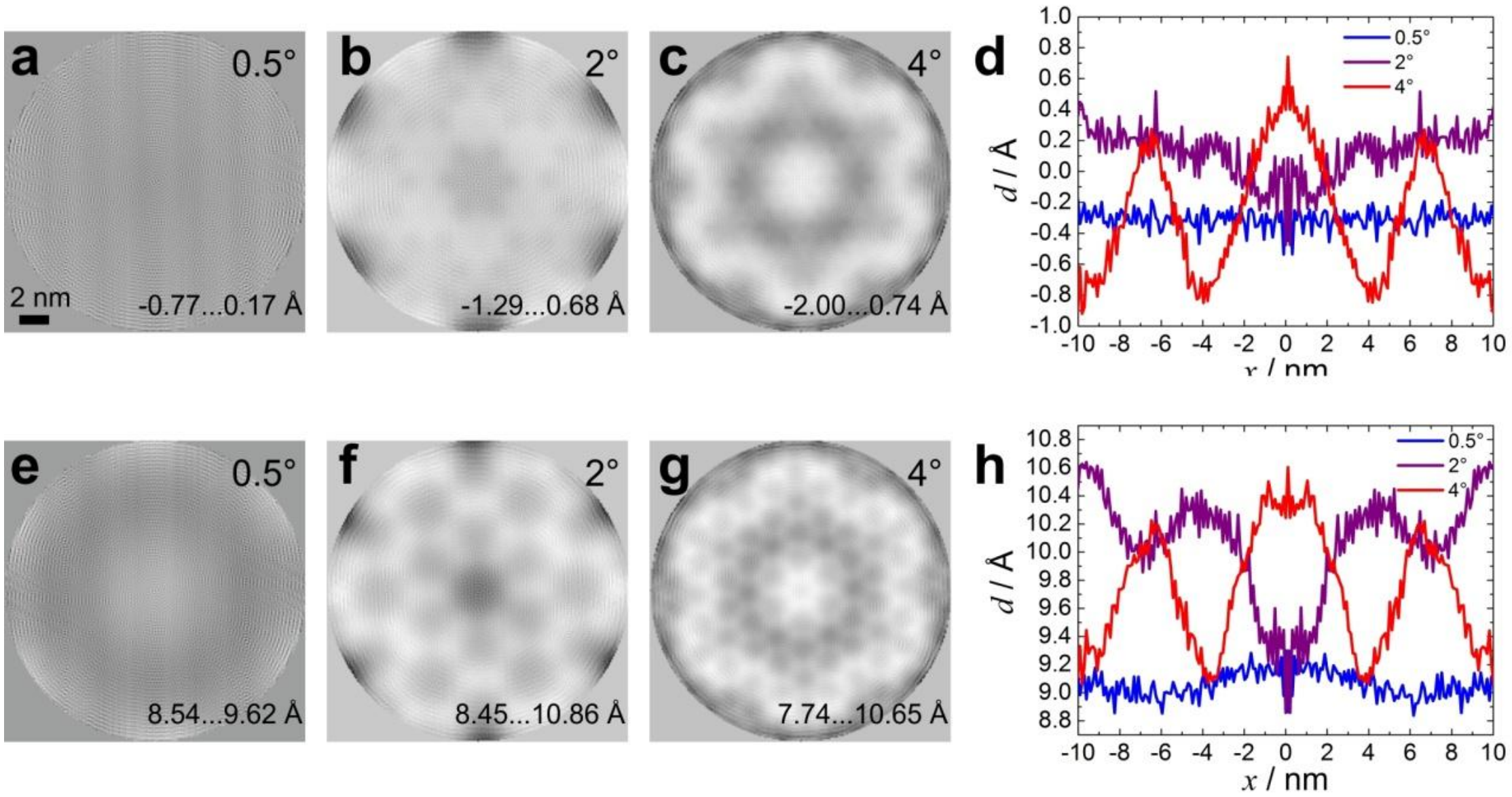

Fig. 6. 2D distributions ((a)-(c) and (e)-(g)) and the profiles through the middle of the 2D distributions ((d) and (h)) of the reconstructed interlayer distances from CBED patterns of BLG with the interlayer distances of 0 (a)-(d) and 10 Å (e)-(h), for three different twist angles of 0.5°, 2° and 4°. Defocus distance is -2 µm.

Figure 6 shows the 2D distributions and profiles of the reconstructed interlayer distances for BLG with interlayer distances of 0 and 10 Å and three different twist angles of 0.5°, 2° and 4°. From Fig. 6, we see that the reconstructions obtained from CBED patterns with smaller twist angles exhibit a smoother appearance, whereas at larger twist angles, some artefact modulations are observed in the reconstructions. These artefacts can be explained by the presence of a moiré structure, which is more apparent at larger twist angles. At larger twist angles, the associated moiré peaks [19] are also observed in the Fourier spectra, as indicated by the red arrows in Fig. 5(c). The precision of the reconstructed interlayer distance is about ±0.5 Å.

We now define the condition at which at least one fringe (one period) appears in a first-order CBED spot. The period is given by Eq. (9) and the CBED spot diameter in *K*-coordinates is given by:

$$D_K = \frac{4\pi}{\lambda} \sin\alpha,$$

where $\alpha$ is the semi-convergence angle. At $T < D_k$ , there is at least one fringe (period) in the CBED spot, which in turn is sufficient for the holographic reconstruction.

### *4.2 Reconstruction of average interlayer distance*

The moiré structure, appearing as an artefact in the reconstructed interlayer distance, can be suppressed if instead of the interlayer distribution over the probed region only an average interlayer distance is reconstructed. This is achieved as follows. During the reconstruction by filtering in the Fourier domain, instead of selecting half of the Fourier spectrum, only one pixel corresponding to the maximum of the sideband is selected and the remaining pixels are set to zero. This single pixel when shifted to the centre will give a constant distribution in real space after the inverse Fourier transform is applied. This can be also considered as an extreme low-pass filter. As a result, the artefact moiré pattern is removed in the reconstructed phase distribution and the interlayer distance distribution is a constant, as shown in Fig. 7.

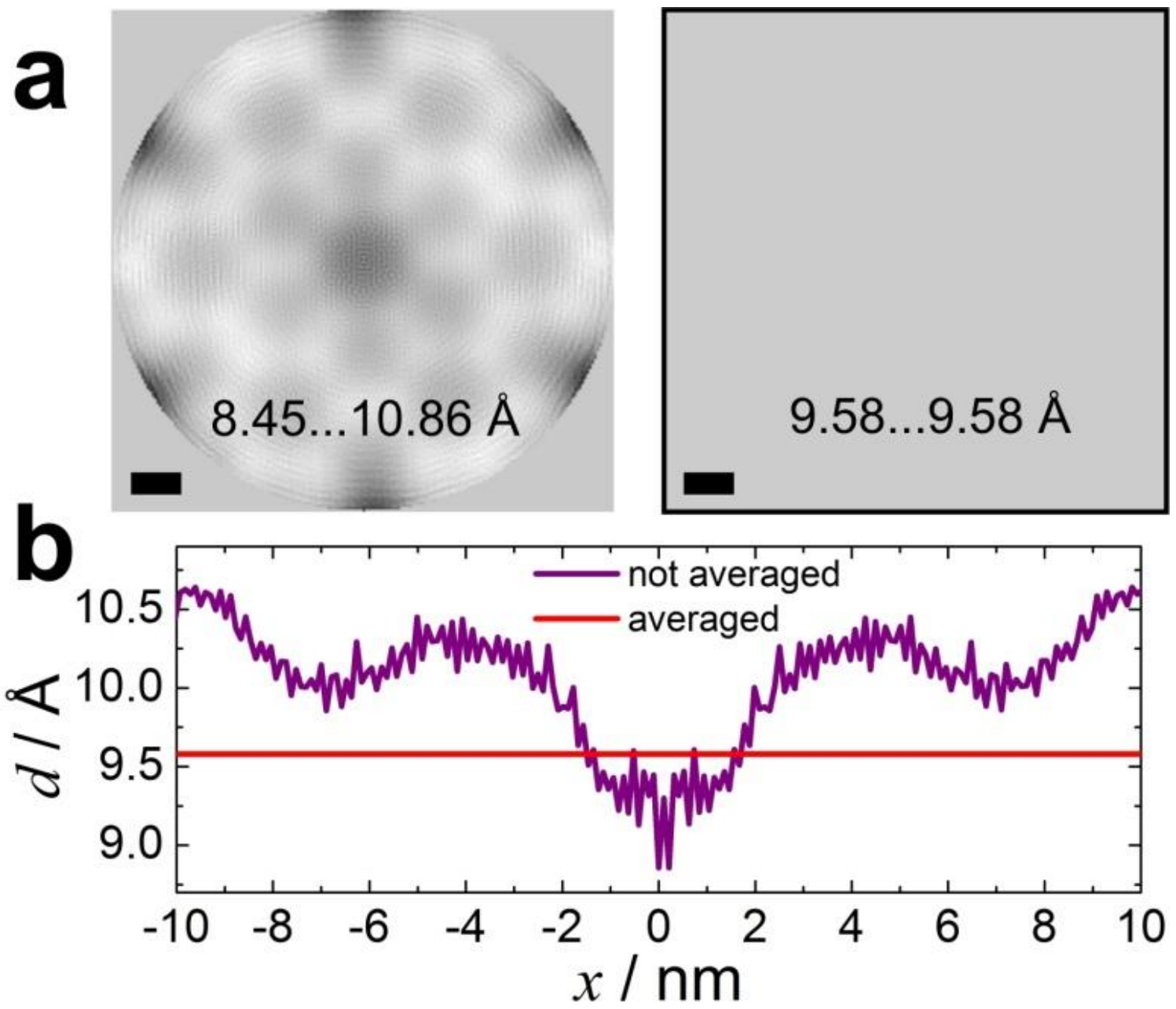

Fig. 7. Reconstruction of average interlayer distance from a CBED pattern of BLG with the interlayer distance of 10 Å, a twist angle of 2° and imaged at the defocus distance of -2 μm. (a) 2D distributions of interlayer distance obtained by selecting half of the spectrum (left) and only maximum of the sideband (right) in the Fourier spectrum. The scalebar is 2 nm. (b) Profiles through the middle of the distributions shown in (a).

### *4.3 Variable interlayer distance*

In reality, the interlayer distance is not a constant but always exhibits some variations. In order to check the sensitivity of the holographic reconstruction method to the interlayer distance variations, we simulated a BLG sample where the distance between the layers is not constant but contains an out-of-plane ripple in the form of a fringe. The distance between the layers is assumed to be 6 Å and atoms in one of the layers are shifted by $\Delta z = -A\exp\left(-\frac{x^2}{2\sigma^2}\right)$, $A = 100$ pm and $\sigma = 2$ nm. The other parameters are: the twist angle is of 0.5° and the defocus distance is -2 $\mu$m. A small twist angle is chosen to minimise the moiré effect. The reconstructed interlayer distribution is shown in Fig. 8. The deviation of the atomic z-position from constant are readily picked up in the interference pattern of overlapping CBED spots and clearly manifests itself in the reconstruction (Fig. 8). However, quantitatively, the recovered $\Delta z$ shifts are much larger than the actual $\Delta z$ shifts. We therefore conclude that even such small variations in the interlayer distance as 0.1 Å will be evident in the reconstructed interlayer distance distribution; however, they will be greatly enhanced and their exact value will be reconstructed with a large error.

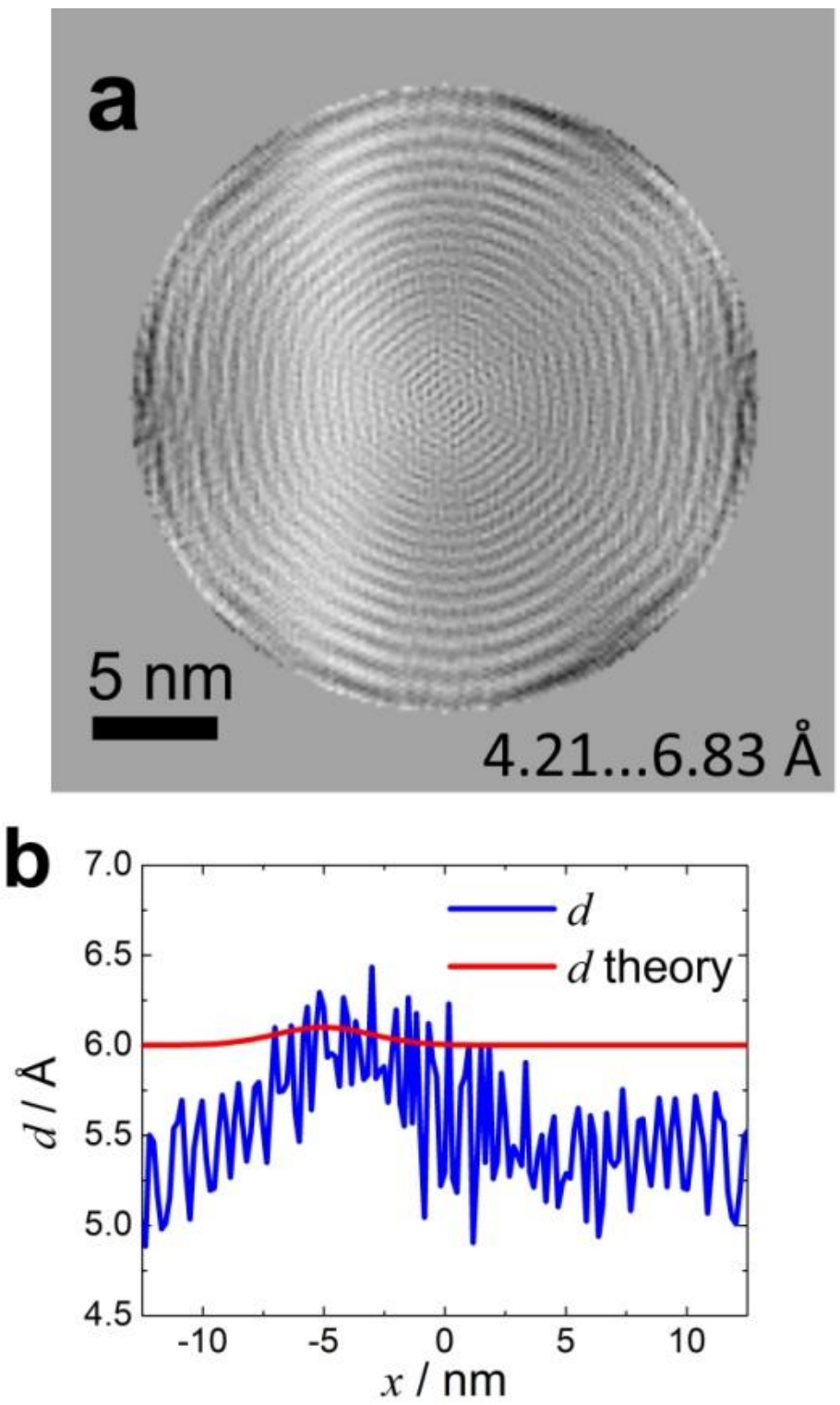

Fig. 8. Reconstruction of interlayer distance from a CBED pattern of BLG with a variable interlayer distance. Atoms in one of the layers were shifted by $\Delta z = A\exp\left(-\frac{x^2}{2\sigma^2}\right)$,

*A* = 100 pm and $\sigma$ = 2 nm. The interlayer distance is *d* = 6 Å, the twist angle is $\varphi$ = 0.5° and imaged at the defocus distance of -2 μm. (a) Reconstructed 2D distributions of the interlayer distance. The scalebar is 5 nm. (b) Profiles through the middle of the distribution shown in (a) with the theoretical distribution of the interlayer distance.

### *4.4 Resolution*

The lateral and axial (along the *z* axis) resolution evaluated from a CBED pattern *k*-value range is given by $d_{x,y} = \frac{\lambda}{\sin \vartheta_{\max}}$ and $d_z = \frac{\lambda}{1 - \cos \vartheta_{\max}}$, respectively, where $\vartheta_{\max}$ is the maximal detected diffraction angle in the CBED pattern, as shown in Fig. 9. According to these equations, for a BLG CBED pattern acquired only up to the first-order CBED spots, the lateral resolution is $d_{x,y} = 2.13$ Å and the axial resolution is $d_z = 217.2$ Å. It is therefore a remarkable result that the holographic approach allows reconstruction of the interlayer distance at 0.5 Å accuracy, which is more than 400 times the diffraction defined *z*-resolution.

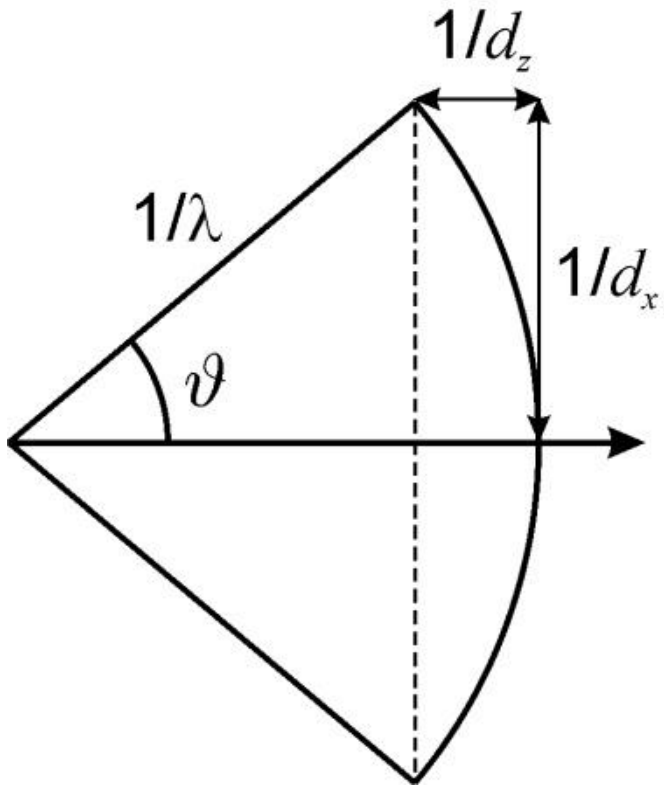

Fig. 9. Illustration of the symbols used for deriving the resolution criteria.

## 5. Conclusions and discussion

We have derived a formula for recovering the interlayer distance from the phase distribution reconstructed from CBED patterns. The formula accounts for the different chemical compositions of the individual monolayers, and is practical for samples such as graphene-BN. We show that surprisingly even one interference fringe in the interference pattern is sufficient to reconstruct the phase shift and with this, the interlayer distance. One interference fringe is observed in CBED spots when either the twist angle or the probed region (defined by the defocus distance) is small. In fact, the situation when only one interference fringe is observed in CBED spots is preferred, as it allows suppression of the artefact signal due to the moiré structure.

The precision of the reconstructed interlayer distance is about ±0.5 Å. It should be noted that this is a few hundred times better than the intrinsic z-resolution evaluated from the spread of the measured *k*-values. Finally, we demonstrated that, in principle, the holographic CBED imaging can pick as small as 0.1 Å variations in the interlayer distance. Though the quantitative reconstruction of such variations suffers from large errors.

## Appendix A: Simulation procedure

The transmission functions of MLs were calculated as:

$$t(x,y) = \exp\left[i\sigma v_z(x,y) \otimes l(x,y)\right], \tag{A1}$$

where $v_z(x,y)$ is the projected potential of an individual atom, $l(x,y)$ is the function describing positions of the atoms in the lattice, and $\otimes$ denotes convolution. The projected potential of a single carbon atom was simulated in the form [20]:

$$v_z(r) = 4\pi^2 a_0 e \sum_i a_i K_0\left(2\pi r\sqrt{b_i}\right) + 2\pi a_0 e \sum_i \frac{c_i}{d_i} \exp\left(-\pi^2 r^2 / d_i\right),$$

where $r = \sqrt{x^2 + y^2}$, $a_0$ is the Bohr' radius, $e$ is the elementary charge, $K_0(\ldots)$ is the modified Bessel function and $a_i, b_i, c_i, d_i$ are parameters that depend on the chemical origin of the atoms and are tabulated in Ref. [20]. In $v_z(r)$, the singularity at $r = 0$ was replaced by the value of $v_z(r)$ at *r* = 0.1 Å. The convolution $v_z(x,y) \otimes l(x,y)$ in Eq. (A1) was calculated as $\mathrm{FT}^{-1}\left\{\mathrm{FT}\left[v_z(x,y)\right]\mathrm{FT}\left[l(x,y)\right]\right\}$, where FT denotes Fourier transform. $\mathrm{FT}\left[l(x,y)\right]$ was simulated without applying Fast Fourier transforms (FFT) to avoid artefacts associated with FFT, it was calculated as $\mathrm{FT}\left[l(x,y)\right] = \sum_n \exp\left[-i\left(k_x x_n + k_y y_n\right)\right]$, where $(x_n, y_n)$ are the exact atomic coordinates of *n*-th atom, not pixels. The inverse FT was calculated by applying inverse FFT to the product of $\mathrm{FT}\left[v_z(x,y)\right]$ and $\mathrm{FT}\left[l(x,y)\right]$.

The incident convergent wave distribution $\psi_0(\vec{r})$ was calculated by simulation diffraction of a spherical wavefront on an aperture (second condenser aperture) positioned at a plane $\vec{r}_0$:

$$\psi_0(\vec{r}) \propto \iint a(\vec{r}_0) \frac{\exp(-ikr_0)}{r_0} \frac{\exp\left(ik\left|\vec{r}_0 - \vec{r}\right|\right)}{\left|\vec{r}_0 - \vec{r}\right|} \mathrm{d}\vec{r}_0,$$

where $a(\vec{r}_0)$ is the aperture function. Each ML was assigned a transmission function $t_i(x_i, y_i)$ defined by Eq. (A1), where $i = 1, 2$ is the layer number. No weak phase object approximation was applied in the simulations. The exit wave after passing through the first layer was given by $u_1(x_1, y_1) = \psi_0(x_1, y_1) t_1(x_1, y_1)$. Next, this wave was propagated to the second layer. The propagation was calculated by the angular spectrum method [20-22]. The distribution of the propagated wave in the plane $(x_2, y_2)$ is given by the complex-valued distribution $u_{2,0}(x_2, y_2)$. The exit wave after passing the second ML was calculated as $u_2(x_2, y_2) = u_{2,0}(x_2, y_2) t_2(x_2, y_2)$. The CBED pattern was then simulated as the square of the amplitude of the FT of $u_2(x_2, y_2)$, where the FT was calculated by FFT. The distributions in the sample is sampled $N \times N$ pixels and the pixel size is $\Delta_0$, meaning the total sample area is thus $N\Delta_0 \times N\Delta_0$. The pixel size in the diffraction plane $\Delta_k = 1/(N\Delta_0)$.

## Acknowledgements

This work was supported by the European Union Graphene Flagship Program; European Research Council Synergy Grant 319277 “Hetero2D” and European Research Council Starting Grant 715502 “EvoluTEM”; the Royal Society; Engineering and Physical Research Council (UK); and US Army Research Office (Grant W911NF-16-1-0279). S.J.H. and E.P. acknowledge funding from the Defense Threat Reduction Agency (Grant HDTRA1- 12-1-0013) and the Engineering and Physical Sciences Research Council (UK) (Grants EP/K016946/1, EP/L01548X/1, EP/M010619/1, and EP/P009050/1).